\documentclass[prb, aps, amsmath, amssymb, amsfonts, twocolumn, superscriptaddress]{revtex4}
\usepackage[german,american,english]{babel}
\usepackage{graphicx}
\usepackage{graphics}
\usepackage{dcolumn}
\usepackage{bm}
\usepackage{amssymb}
\usepackage{amsmath}
\usepackage{amsfonts}
\usepackage{epsfig}

\newcommand {\el} {\\ \nonumber}

\newcommand {\ket} [1] {| #1 \rangle}
\newcommand {\bkt} [1] {\langle #1 \rangle}

\newcommand {\pd} [2] {\frac{\partial #1}{\partial #2}}

 \newcommand {\beq}{\begin{equation}}
\newcommand {\eeq}{\end{equation}}
\newcommand {\beqn}{\begin{eqnarray}}
\newcommand {\eeqn}{\end{eqnarray}}
\newcommand {\bit}{\begin{itemize}}
\newcommand {\eit}{\end{itemize}}
\newcommand {\br}{\langle}
\newcommand {\ke}{\rangle}

\newcommand {\ee}{\varepsilon}

\newcommand {\bk}{\mathbf{k}}

\newif\ifContLineOne
\newif\ifContLineTwo
\newif\ifContLineThree

\def\conC#1{\vbox{\ialign{##\crcr
  \ifContLineThree\hrulefill\else\vphantom{\hrulefill}\fi\crcr
  \noalign{\kern3.2pt\nointerlineskip}
  \ifContLineTwo\hrulefill\else\vphantom{\hrulefill}\fi\crcr
  \noalign{\kern3.2pt\nointerlineskip}
  \ifContLineOne\hrulefill\else\vphantom{\hrulefill}\fi\crcr
  \noalign{\nointerlineskip}
  $\hfil\textstyle{\vbox to 14pt{}#1}\hfil$\crcr}}}

\def\DrawLeg#1#2{
  \kern-.2pt              
  \dimen2 =#1             
  \advance\dimen2 by 2pt  
  \dimen3 = 10.6pt        
  \dimen4 =3.6pt          
  \advance\dimen3 by -\dimen2
  \multiply\dimen4 by #2
  \advance\dimen3 by \dimen4
  \raise\dimen2 \hbox{\vrule height\dimen3 width .4pt} 
  \kern-.2pt}             

\def\begC#1#2{\setbox0 =\hbox{$\textstyle{#2}$}
  \dimen0=.5\wd0 \dimen1=\ht0
  \conC{\hskip\dimen0}
  \count255=#1
  \ifnum\count255 =1 \ContLineOnetrue\else
  \ifnum\count255 =2 \ContLineTwotrue\else
  \ifnum\count255 =3 \ContLineThreetrue\fi\fi\fi
  \DrawLeg{\dimen1}{\count255}
  \conC{\hskip\dimen0}
  \kern-\dimen0\kern-\dimen0 \box0}

\def\endC#1#2{\setbox0 =\hbox{$\textstyle{#2}$}
  \dimen0=.5\wd0 \dimen1=\ht0
  \conC{\hskip\dimen0}
  \count255=#1
  \ifnum\count255 =1 \ContLineOnefalse\else
  \ifnum\count255 =2 \ContLineTwofalse\else
  \ifnum\count255 =3 \ContLineThreefalse\fi\fi\fi
  \DrawLeg{\dimen1}{\count255}
  \conC{\hskip\dimen0}
  \kern-\dimen0\kern-\dimen0 \box0}

\begin{document}
\title{Suppression of the Kondo Resistivity Minimum in Topological Insulators}
\author{Jie Wang}
\affiliation{ICQD, Hefei National Laboratory for Physical Sciences at the Microscale, University of Science and Technology of China, Hefei 230026, Anhui, China}
\author{Dimitrie Culcer}
\affiliation{School of Physics, The University of New South Wales, Sydney 2052, Australia}
\affiliation{ICQD, Hefei National Laboratory for Physical Sciences at the Microscale, University of Science and Technology of China, Hefei 230026, Anhui, China}
\begin{abstract}
Magnetically-doped topological insulators are intensely studied in the search for exotic phenomena such as the quantum anomalous Hall effect. The interplay of electronic and impurity degrees of freedom leads to the Kondo effect, an increase in the resistivity at temperatures $T < T_K$, the Kondo temperature. We study this effect in chiral surface state transport at $T \ge T_K$ in the metallic regime, starting from the quantum Liouville equation and including Kondo scattering to all orders, as well as phonon and non-magnetic impurity scattering. Unlike spin-orbit coupled metals and semiconductors, $T_K$ is suppressed by spin-momentum locking which prevents the formation of a Kondo screening cloud. We expect a resistivity $\rho_{xx} \propto T^4$ primarily due to phonons.
\end{abstract}
\date{\today}
\maketitle

\section{Introduction\label{Sec:Intro}}
Considerable attention has been devoted lately to topological insulators (TI), which have an insulating bulk and conducting edge (2D) or surface (3D) states protected by time-reversal symmetry \cite{Hasan_TI_RMP10, Qi_TI_RMP_10}. Strong spin-orbit coupling in TI leads to a dispersion in the form of a Dirac cone, suppressed backscattering, and coupled charge and spin transport \cite{Tkachov_TI_Review_PSS13, Culcer_TI_PhysE12}. Within this field, TI doped with magnetic impurities have been the focus of an intense effort, culminating in the observation of the quantum anomalous Hall effect \cite{Yu_TI_QuantAHE_Science10, QAHE}.

Magnetically-doped systems frequently exhibit an increase in the the resistivity below a certain \textit{Kondo temperature} $T_K$. The Kondo effect stems from the interplay between electron and impurity spins resulting in spin-flip scattering, which leads to screening of the impurity spin below $T_K$ \cite{Kondo_64,Anderson_PR61, Hewson}. In systems with spin-orbit coupling this effect is of considerable interest, given the associated spin non-conservation and nontrivial spin dynamics. Studies have focused on spin-orbit coupled semiconductors \cite{Meir_PRB1994, Malecki_JSP07, Zarea_PRL12, Zitko_PRB11, Feng_RashbaDres_Kondo_JPCM11}, including quantum dots \cite{Kikoin_PRB2012}, non-centrosymmetric metals \cite{Vekhter_PRB12}, and superconductors \cite{Yanagisawa_JPSJ12}. In particular, Refs.~\onlinecite{Zarea_PRL12, Vekhter_PRB12} showed that spin-orbit coupling can enhance the Kondo temperature.

In this context, TI are especially interesting, since the spin-orbit interaction is the dominant energy scale. TI Kondo physics is conceptually unique due to the interplay of impurity degrees of freedom with the spin-momentum locking of the conduction electrons, offering an example of the competition between strong spin-orbit coupling and strong interactions (these systems are distinct from topological Kondo insulators \cite{Dzero_TKI_PRL10}). Research on the Kondo effect in 2DTI \cite{Maciejko_2DTI_Kondo_PRL09, Maciejko_2DTI_KondoLtc_PRB12} and 3DTI \cite{Biswas_TI_ImpurStt_PRB10, Tran_PRB10, Zitko_TI_Kondo_PRB10, Feng_TI_Kondo_PRB10, Ng_TI_QmImp_PRB13, Ng_TI_AndersonImp_RG_13, Fritz_Gfn_Kondo_RPP13, Mitchell_TI_Kondo_PRB13} is taking off. Studies to date have largely focused on low-temperatures and doping near the Dirac point, with Ref.~\onlinecite{Zitko_TI_Kondo_PRB10} mapping the problem onto the Anderson pseudogap model. Spin-orbit coupling gives a strong anisotropy in the correlation of the impurity and conduction electron spin densities \cite{Feng_TI_Kondo_PRB10}, and a universal energy dependence of the low-energy quasiparticle interference near the Dirac point \cite{Mitchell_TI_Kondo_PRB13}. Interestingly, the Kondo resonance in the \textit{bulk} of TI can be screened by the exchange interaction \cite{Ng_TI_QmImp_PRB13, Ng_TI_AndersonImp_RG_13}.

Fundamental questions remain, especially in regard to the role of spin-momentum locking in the Kondo effect in 3DTI transport. Because of spin-momentum locking, momentum scattering in TI is always accompanied by spin rotations, meaning that one cannot simply translate results known for dilute alloys \cite{Nagaoka_Kondo_PR65, Hamann_Kondo_EqMtn_PRB67, Silverstein_Duke_PhysRev.161.470, AbrikosovMigdal_Kondo_JLTP70}. However, despite similarities with graphene \cite{Tran_PRB10}, the TI Hamiltonian describes the true spin in a one-valley system \cite{Fritz_Gfn_Kondo_RPP13}. This makes large-$N$ renormalization a difficult proposition in TI, given that $N = 1$. Moreover, the topological protection offered by suppressed backscattering is not meaningful out of equilibrium \cite{Culcer_TI_Kineq_PRB10}, since transport is irreversible. At the same time, the observation of chiral surface states in transport has been problematic \cite{Culcer_TI_PhysE12}, and recent experiments have only isolated their contribution by using gates \cite{Kim_TI_Gate_MinCond_NP12, Kim_TI_e-ph_PRL12, Kim_TI_CohTnsp_NC13}. Given the current low sample qualities, it is essential to characterize the surface states fully and identify transport signatures, including $T_K$, and the location of the resistance minimum, which in general occurs at a temperature different from $T_K$, requiring the full resistivity for its evaluation, including the phonon contribution.

In light of this, we present here a transport theory of non-equilibrium magnetic 3DTI that treats impurity, phonon \cite{Qiuzi_TI_Tnsp_PRB12} and Kondo scattering on the same footing. Since gating can eliminate bulk transport \cite{Kim_TI_Gate_MinCond_NP12, Kim_TI_e-ph_PRL12, Kim_TI_CohTnsp_NC13}, we focus on the surface states alone. We derive a many-body kinetic equation from the quantum Liouville equation and sum the scattering terms to all orders in the Kondo interaction, retaining the leading divergent terms, the equivalent of the parquet diagrams. We derive the resistivity as a function of $T$, showing that $T_K$ is strongly suppressed, and the temperature dependence of the resistivity is primarily due to phonon scattering. Physically, this is because spin momentum locking makes it difficult for the impurity spin to polarize the conduction electrons.

The outline of this paper is as follows. In Sec. \ref{Sec:Hamiltonian}, we will introduce the Hamiltonian of the system, including the band, impurity and driving electric field contributions. Section \ref{Sec:Trans} focuses on the transport theory of magnetically doped TI, deriving the kinetic equation directly from the quantum Liouville equation. The full resistivity and Kondo temperature are also derived in this section. The results are discussed in Sec.~\ref{Sec:Disc}. Finally, Sec.~\ref{Sec:Summary} summarizes our findings.

\section{Hamiltonian}
\label{Sec:Hamiltonian}

We focus on temperatures $T\geq T_K$ and assume $\ee_F\tau/\hbar>>1$, where $\ee_F$ is the Fermi energy and $\tau$ the momentum relaxation time, and $\varepsilon_F$ lies in the bulk gap but in the surface conduction band. Single electron states $\ket{{\bm k}s}$ below are indexed by wave vector ${\bm k}$ and spin $s$. The total Hamiltonian is $H = H_0 + U$, where $H_0 = H_{TI} + H_E$ and the total effective scattering potential $U = U_{imp} + U_{m} + U_{ep}$. The surface state band Hamiltonian is $H_{TI} = \displaystyle -\sum_{{\bm k}ss'} Ak{\bm \sigma} ^{ss'}\cdot \hat{\bm \theta} \, c^{\dag}_{{\bm k} s} c_{{\bm k}s'}$, where ${\bm \sigma}$ is the electron spin operator, $\hat{\bm \theta}$ is the tangential unit vector corresponding to ${\bm k}$, and $A$ is a constant. The interaction with the driving electric field, $H_E = \displaystyle \sum_{{\bm k}{\bm k}'ss'} H^E_{{\bm k}{\bm k}'} c^{\dag}_{{\bm k} s} c_{{\bm k}'s'}$, with $H^E_{{\bm k}{\bm k}'}$ given below.

The potential due to non-magnetic charged impurities and static defects is $U_{imp} = \displaystyle \sum_{I{\bm k}{\bm k}'s} V^C_{{\bm k}{\bm k}'} c^{\dag}_{{\bm k} s} c_{{\bm k}s}$, where $V^C_{{\bm k}{\bm k}'} = \displaystyle \bar{V}^C_{{\bm k}{\bm k}'} \sum_{J} e^{-i({\bm k} - {\bm k}')\cdot{\bm R}_J}$. Here $\bar{V}^C_{{\bm k}{\bm k}'}$ is the Coulomb potential of a single impurity and ${\bm R}_J$ denotes the impurity locations. The impurities are assumed uncorrelated and the average of $V^C_{{\bm k}{\bm k}'}V^C_{{\bm k}'{\bm k}}$ over impurity configurations is $(n_i |\bar{V}^C_{{\bm k}{\bm k}'}|^2 \delta_{ss'})/V$, where $n_i$ is the impurity density and $V$ the crystal volume. Scattering is assumed elastic.

The Kondo interaction $U_{m} = \displaystyle \sum_{I{\bm k}, {\bm k}', s, s'} W_{I{\bm k}{\bm k}'} ^{ss'} \, c^{\dag}_{{\bm k} s} c_{{\bm k}'s'}$, where $ W_{I{\bm k}{\bm k}'}^{ss'} = \displaystyle \bar{W}_{I{\bm k}{\bm k}'}^{ss'} e^{-i({\bm k} - {\bm k}')\cdot{\bm R}_I}$, describes scattering off \textit{magnetic} impurities with density $n_m$, assumed local in space, and $I$ runs over magnetic impurities. For a single impurity $\bar{W}_{I{\bm k}{\bm k}'}^{ss'} = (J/V) {\bm \sigma}^{ss'} \cdot {\bm S}^I$ or
\begin{equation}
\arraycolsep 0.3ex
\begin{array}{rl}
\displaystyle \bar{W}_{I{\bm k}{\bm k}'}^{ss'} = \frac{J}{V}[\sigma_zS_z^I+\frac{1}{2} (S^I_+\sigma_-+ S^I_-\sigma^+)]^{ss'},
\end{array}
\end{equation}
where $S_I$ are impurity spin operators, and $\sigma_\pm = \sigma_x \pm i \sigma_y$.

The electron-phonon interaction is
\begin{equation}
U_{ep}=\sum_{\mathbf{k},\mathbf{q},s}D_{q}c_{\mathbf{k}+\mathbf{q},s}^{\dag}c_{\mathbf{k},s}(b_{\mathbf{q}}+b^{\dag}_{-\mathbf{q}}),
\end{equation}
with $b_{\mathbf{q}}$, $b_{\mathbf{q}}^{\dag}$ phonon annihilation/creation operators, the deformation potential $D_{q} = - iC\sqrt{\frac{\hbar q}{2\rho v_{ph}}}$, $C\approx 30$ eV\cite{Qiuzi_TI_Tnsp_PRB12}, $\rho$ the mass density, and $v_{ph}$ the phonon velocity.

\subsection{Eigenstates of TI Hamiltonian}
The eigenstates $|\bk n\ke$ of $H_{TI}$ are denoted by $|\bk,\pm\ke$, where $\pm$ represent the electron and hole bands respectively
\beqn
|\bk,+\ke&=&\frac{1}{\sqrt{2}}[e^{-i\theta/2}|\bk,\uparrow\ke-ie^{i\theta/2}|\bk,\downarrow\ke]\el
|\bk,-\ke&=&\frac{1}{\sqrt{2}}[e^{-i\theta/2}|\bk,\uparrow\ke+ie^{i\theta/2}|\bk,\downarrow\ke].
\eeqn
Matrix elements in the eigenstate basis carry a tilde $\tilde{}$. The $x$-velocity operator in the eigenstate basis is
\beqn
\tilde{v}_x&=&\frac{1}{\hbar}\pd{H}{k_x} = \frac{A\cos\theta}{\hbar} \, \sigma_{k||} + \frac{A\sin\theta}{\hbar} \, \sigma_{k\bot}.
\eeqn

A screened Coulomb potential $\bar{V}^C$ in the Pauli basis $\ket{{\bm k}s}$ is $\bar{V}_{\bk\bk'}^{C} = \frac{Ze^2}{2\ee_0\ee_r}\frac{1}{|\bk - \bk'|+k_{TF}} \, \openone$, where $k_{TF}$ is the Tomas-Fermi wave vector \cite{Culcer_TI_PhysE12} and $\openone$ the identity matrix in spin space. In the basis of TI eigenstates
\beqn
\tilde{V}^C_{\bk\bk'} & = & S_{\bk}^{\dag}\bar{V}^{C}_{\bk\bk'}S_{\bk'} = \frac{Ze^2}{2\ee_0\ee_r}\frac{1}{|\bk - \bk'|+k_{TF}} \,
\left(
\begin{array}{cccc}
\cos\frac{\gamma}{2}&-i\sin\frac{\gamma}{2}\\
-i\sin\frac{\gamma}{2}&\cos\frac{\gamma}{2}
\end{array}\right)
\eeqn
where $\gamma=\theta' - \theta$ and the transfer matrix
\beqn
S_{\bk}&=&\left(
\begin{array}{cccc}
\br \bk,\uparrow|\bk,+\ke&\br \bk,\uparrow|\bk,-\ke\\
\br \bk,\downarrow|\bk,+\ke&\br \bk,\downarrow|\bk,-\ke
\end{array}\right) = \frac{1}{\sqrt{2}}\left(
\begin{array}{cccc}
e^{-i\theta/2}&e^{-i\theta/2}\\
-ie^{i\theta/2}&ie^{i\theta/2}
\end{array}\right).
\eeqn
The Kondo interaction with a single magnetic impurity has the following matrix elements in eigenstate space
\beqn\label{Wkk'}
\tilde{W}_{\bk\bk'}^{++} & = & \frac{J}{2} \, (-2i\sin\frac{\gamma}{2}S^z+ie^{-i\phi/2}S^+-ie^{i\phi/2}S^-)\el
\tilde{W}_{\bk\bk'}^{+-} & = & \frac{J}{2} \, (2\cos\frac{\gamma}{2}S^z+ie^{-i\phi/2}S^++ie^{i\phi/2}S^-)\el
\tilde{W}_{\bk\bk'}^{-+} & = & \frac{J}{2} \, (2\cos\frac{\gamma}{2}S^z-ie^{-i\phi/2}S^+-ie^{i\phi/2}S^-)\el
\tilde{W}_{\bk\bk'}^{--} & = & \frac{J}{2} \, (-2i\sin\frac{\gamma}{2}S^z-ie^{-i\phi/2}S^++ie^{i\phi/2}S^-),
\eeqn
where $\phi=\theta+\theta'$. Notice the presence of backscattering terms. We remark in addition that $\br \tilde{W}_{\alpha\beta}\tilde{W}_{\beta\gamma}\tilde{W}_{\gamma\alpha}\ke=\br \tilde{W}_{\beta\gamma}\tilde{W}_{\gamma\alpha}\tilde{W}_{\alpha\beta}\ke$, $\br \tilde{W}_{\alpha\beta}\tilde{W}_{\beta\gamma}\tilde{W}_{\gamma\alpha}\ke=-\br \tilde{W}_{\alpha\beta}\tilde{W}_{\gamma\alpha}\tilde{W}_{\beta\gamma}\ke$.



\section{Kinetic equation}
\label{Sec:Trans}

The system is described by the many-body density operator $F$. The single-particle density matrix $f_{\alpha\beta} = {\rm Tr} \, (Fc_{\beta}^{\dag}c_{\alpha})$, where $\ket{\alpha} \equiv \ket{\bk_{\alpha}s_{\alpha}}$ and Tr is the full operator trace. $F$ obeys the quantum Liouville equation
\beqn \label{QLE}
\frac{d F(t)}{dt}+\frac{i}{\hbar}[H, F(t)] = 0.
\eeqn
Assuming random impurity locations and spins $R_I$, $m$, we introduce the impurity average of $F$ through
\beqn
\br F(t)\ke=\Pi_{I=1}^{n}\int \frac{dR_I}{V} \frac{1}{2S+1} \, \sum_{m=-S}^{S} \br m|F(R_1,...R_n; m; t)|m\ke\label{average}.
\eeqn
We write $F = \bkt{F} + G$, where $\bkt{F}$ is averaged over impurities and $G = F - \br F \ke$ is the remainder. We integrate over $G$, since our interest is in impurity-averaged expectation values, hence $\bkt{F}$. Then Eq.\ (\ref{QLE}) is broken up into
\beqn
\frac{d\br F(t)\ke}{dt}+\frac{i}{\hbar}[H_0,\br F(t)\ke]  \ke & = & - \frac{i}{\hbar}\br[U,G(t)] \label{quantum liouville equation}\el
\frac{dG(t)}{dt}+\frac{i}{\hbar}[H_0,G(t)]+\frac{i}{\hbar}[U,\br F(t)\ke] & = & \frac{i}{\hbar}\bkt{[U,G(t)]}.
\eeqn

The scattering term is $\mathcal{J}(F) = \displaystyle (i/\hbar) \br[U,G(t)]\ke$. We solve
\beqn
G(t)= - \frac{i}{\hbar}\int_0^{\infty}dt'e^{-iHt'/\hbar}[U,\br F(t-t')\ke]e^{iHt'/\hbar},
\eeqn
and introduce resolvents $R^{\pm}(E) = (E-H\pm i\eta)^{-1}$ in Fourier space, with $\eta$ infinitesimal. The resolvents satisfy
\beqn
e^{\mp iHt/\hbar}e^{-\eta t}&=&\pm\frac{i}{2\pi}\int_{-\infty}^{\infty}dER^{\pm}(E)e^{\mp iEt/\hbar}\label{r1}\\
R^{\pm}(E)&=&\frac{1}{\pm i\hbar}\int_0^{\infty}dte^{\mp iHt/\hbar}e^{\pm iEt/\hbar}e^{-\eta t}\label{r2}
\eeqn
We also introduce the $T$ operators, given by $T^{\pm}(E) = U + UR^{\pm}(E)U$. Finally, we obtain
\beqn
\mathcal{J} &=& -\int_{-\infty}^{\infty}\frac{dE}{2\pi\hbar}\br A(E,t)\ke+h.c.
\eeqn
where $h.c.$ stands for Hermitian conjugate, the function
\begin{widetext}
\beqn
A(E,t) = T^+(E)[R_0^+(E)\br F(t)\ke-\br F(t)\ke R_0^-(E)]T^-(E)R_0^-(E)+R_0^+(E)\br F(t)\ke T^+(E)[R_0^+(E)-R_0^-(E)]T^-(E)\label{A},
\eeqn
\end{widetext}
and the bare resolvent $R_0^{\pm}(E) = (E - H_{TI} \pm i\eta)^{-1}$.

We use Wick's theorem to obtain a one-particle equation for $f$. We switch to the eigenstate representation $\ket{\gamma} \equiv \ket{\bk n}$, where $n$ is used exclusively for the band index. We focus on the intraband part of $f$, diagonal in $n$, since interband matrix elements are next-to-leading order in $\hbar/\ee_F\tau \ll 1$ \cite{Culcer_TI_Kineq_PRB10}. The equation for $f$ is found by tracing (\ref{quantum liouville equation}) with $c_{\gamma}^{\dag}c_{\gamma}$, hence $\mathcal{J}(f_{\gamma}) = {\rm Tr} \, [\mathcal{J}(F)c_{\gamma}^{\dag}c_{\gamma}]$.

The electric field, assumed constant and uniform, enters through $H^E_{{\bm k}{\bm k}'} = e\mathbf{E}\cdot {\bm r}_{{\bm k}{\bm k}'}$. To linear order in the electric field, $f_{{\bk}n} = f_{0\bk n} + f_{E\bk n}$, where $f_{0{\bm k}n}$ is the equilibrium Fermi-Dirac distribution function for band $n$.

\subsubsection{Born approximation}

We assume no correlations between different scattering mechanisms, so $\bkt{U_{imp} U_{m}}$ = 0, and similarly for all cross terms. The scattering term in the Born approximation is obtained after replacing $T$ matrix in $A(E,t)$ with $U$, which is $A(E,t)+h.c.=UR^+_0[F,U]R^-_0+h.c$. The reduced scattering term is
\beqn\label{Born}
\mathcal{J}_{\gamma}^{(2)}&=&-\frac{i}{\hbar}\sum_{\alpha\beta}\frac{U_{\alpha\beta}U_{\eta\tau}}{\ee_{\alpha}-\ee_{\beta}+i\eta}\br [ c_{\eta}^{\dag}c_{\tau},c_{\gamma}^{\dag}c_{\gamma}c_{\alpha}^{\dag}c_{\beta}]\ke+h.c.\label{second order}
\el &=& \frac{2\pi}{\hbar}\delta(\ee_{\tau}-\ee_{\gamma})\br U_{\gamma\tau}U_{\tau\gamma}\ke(f_{\gamma}-f_{\tau})
\eeqn
In the last step of deriving the above equation, we used Wick's theorem to approximate the statistical average of a series of operators as the sum of their pairings $\begC1{c_{\alpha}^{\dag}}\endC1{c_{\beta}}=\delta_{\alpha\beta}f_{\alpha}$, for example $tr\br F\ke c_{\alpha}^{\dag}c_{\beta}c_{\eta}^{\dag}c_{\tau}=\begC1{c_{\alpha}^{\dag}}\endC1{c_{\beta}}\begC1{c_{\eta}^{\dag}}\endC1{c_{\tau}}+\begC1{c_{\alpha}^{\dag}}\begC2{c_{\beta}}\endC2{c_{\eta}^{\dag}}\endC1{c_{\tau}}$. We also used the property $\br U_{\alpha\beta}U_{\beta\alpha}\ke=\br U_{\beta\alpha}U_{\alpha\beta}\ke$, the averages of any two operators $\hat{A}$ and $\hat{B}$ commute, which is manifest according to the definition of the averaging process (\ref{average}). Analogous approximations are used in higher orders in $U$.

Based on Eq.\ (\ref{Born}), the non-magnetic impurity scattering term in the Born approximation
\beqn
\mathcal{J}_{imp}(f_{\bk +}) = \frac{n_i k_F}{2 A\hbar}\int \frac{d\gamma}{2\pi}|\bar{V}^{C}_{\bk\bk'}|^2(f_{\bk+} - f_{\bk'+})(1+\cos\gamma)
\eeqn
The magnetic impurity scattering term in the Born approximation [note $\tilde{W}^{++}_{\bk\bk'}$ has the angular structure of Eq.\ \ref{Wkk'}]
\beqn
\mathcal{J}_{m}^{(2)}(f_{\bk+}) &=& \frac{n_m k_F}{A \hbar}\int\frac{d\gamma}{2\pi}\br \tilde{W}^{++}_{\bk\bk'}\tilde{W}^{++}_{\bk'\bk}\ke(f_{\bk+} - f_{\bk'+}).
\eeqn
For the electron-phonon interaction in the Born approximation
\begin{widetext}
\beqn
\mathcal{J}_{ep}(f_{{\bk +}}) = &-& \frac{2\pi}{\hbar} \sum_{\bm q} |D_{\bm q}|^2 \delta(\hbar\omega_{\bm q} + \ee_{{\bm k} - {\bm q}, +}-\ee_{{\bk +}})\label{electron-phonon}[N_{\bm q}f_{{\bm k} - {\bm q}, +}(1-f_{{\bk +}}) - (1 + N_{\bm q}) f_{{\bm k}+}(1 - f_{{\bm k} - {\bm q}, +})] \el
&-& \frac{2\pi}{\hbar} \sum_{\bm q}|D_{\bm q}|^2 \delta(\hbar\omega_{\bm q} + \ee_{{\bm k}+} - \ee_{{\bm k} + {\bm q}, +}) [(1 + N_{\bm q})f_{{\bm k} + {\bm q}, +}(1 - f_{{\bm k}, +})-N_{\bm q}f_{{\bm k}+}(1 - f_{{\bm k} + {\bm q}, +})],
\eeqn
\end{widetext}
where $\begC1{b_{\bm q}^{\dag}}\endC1{b_{{\bm q}'}}=N_{\bm q}\delta_{{\bm Q}{\bm q}'}$, and $N_{\bm q}$ is the phonon distribution. We assume the phonons are in equilibrium $N_{\bm q} = 1/[e^{\hbar\omega_q/k_BT}-1]$, which at low $T$ decays exponentially as a function of energy. Therefore we only consider low energy phonons $\hbar\omega_q<<\ee_F$, for which $1/\sqrt{1-q^2/(2k_F)^2}\approx1$ and $\ee_{k} \approx \ee_{k\pm q} \approx \ee_F$, since transport takes place on the Fermi surface.

Below we will assume $\displaystyle f_{E{\bk+}} \propto (e\mathbf{E}\cdot\hat{\mathbf{k}}/\hbar)(\partial f_{0{\bm k}+}/\partial k)$. The non-magnetic impurity scattering term reduces to
\beqn
\mathcal{J}_{imp}(f_{E{\bk +}})=\frac{n_i k_F f_{E{\bk+}}}{4 A\hbar}\int \frac{d\gamma}{2\pi} |\bar{V}^{C}_{\bk\bk'}|^2 (1+\cos\gamma)
\eeqn
The magnetic impurity scattering term is
\beqn
\mathcal{J}_{m}^{(2)}(f_{E{\bk+}}) = \frac{7 n_m J^2 k_F}{12 A\hbar} \, S(S+1)\, f_{E{\bm k}+}.
\eeqn
For acoustic phonons, with $\omega_q \propto q$, and the maximum $q \rightarrow \infty$, the phonon scattering term is
\begin{equation}
\mathcal{J}_{ep}(f_{E{\bm k}+}) = \bigg(\frac{2\pi^5 C^2}{15 Ak_F^2 \rho v_{ph}^5}\bigg) \bigg(\frac{k_BT}{\hbar}\bigg)^4 f_{E{\bm k}+}.
\end{equation}

\subsubsection{Third and higher orders in $U$}

In the third-order scattering term in the Kondo interaction
\beqn
&&\mathcal{J}_m^{(3)}(f_{{\bm k}+})\label{JKondo}\el
 &=& \frac{8\pi}{\hbar} \sum_{{\bm k}_1{\bm k}_2} \br \tilde{W}_{{\bm k}{\bm k}_1}^{++}\tilde{W}_{{\bm k}_1{\bm k}_2}^{++}\tilde{W}_{{\bm k}_2{\bm k}}^{++}\ke f_{{\bm k}_1+}(f_{{\bm k}_2+} - f_{{\bm k}+})\frac{\delta(\ee_{{\bm k}+}-\ee_{{\bm k}_2+})}{\ee_{{\bm k}+} - \ee_{{\bm k}_1+}},
\eeqn
we substitute $f_{{\bk+}} = f_{0{\bk+}} + f_{E{\bk+}}$, since the scattering term involves both the equilibrium density matrix and its correction linear in the electric field. We make use of integrals of the type
\beqn
\int_0^{\infty}\frac{k_{1}dk_{1}}{k - k_1}f_{0k_1+}=\int_0^{\infty}dk_{1}\pd{f_{0k_1+}}{k_1}\bigg(k_1 + k \ln \bigg|\frac{k_1 - k}{k}\bigg|\bigg).
\eeqn
which also occur in higher orders in $U$. Substituting $\displaystyle f_{E{\bk+}} \propto (e\mathbf{E}\cdot\hat{\mathbf{k}}/\hbar)(\partial f_{0{\bm k}+}/\partial k)$ and performing the impurity spin averages we obtain $\displaystyle\bigg(\frac{1}{\tau_m}\bigg)^{(3)}$ as will show below.

The fourth-order scattering term has the general form
\begin{widetext}
\beqn
\mathcal{J}_m^{(4)} (f_{{\bm k}+})&=&\frac{4\pi}{\hbar}\sum_{{\bm k}_1{\bm k}_2{\bm k}_3}\bkt{2\tilde{W}_{{\bm k}_1{\bm k}}^{++}\tilde{W}_{{\bm k}_2{\bm k}_1}^{++}\tilde{W}_{{\bm k}_3{\bm k}_2}^{++}\tilde{W}_{{\bm k}{\bm k}_3}^{++}-2\tilde{W}_{{\bm k}_1{\bm k}}^{++}\tilde{W}_{{\bm k}_2{\bm k}_1}^{++}\tilde{W}_{{\bm k}{\bm k}_3}^{++}\tilde{W}_{{\bm k}_3{\bm k}_2}^{++}\el
&+&\tilde{W}_{{\bm k}{\bm k}_1}^{++}\tilde{W}_{{\bm k}_1{\bm k}_2}^{++}\tilde{W}_{{\bm k}_2{\bm k}_3}^{++}\tilde{W}_{{\bm k}_3{\bm k}}^{++}-\tilde{W}_{{\bm k}_2{\bm k}_1}^{++}\tilde{W}_{{\bm k}_1{\bm k}}^{++}\tilde{W}_{{\bm k}_3{\bm k}_2}^{++}\tilde{W}_{{\bm k}{\bm k}_3}^{++}} \times\frac{f_{{\bm k}_1+}}{\ee_{\bm k}-\ee_{{\bm k}_1}}\frac{f_{{\bm k}_3+}}{\ee_{\bm k}-\ee_{{\bm k}_3}}(f_{\bm k}-f_{{\bm k}_2})\delta(\ee_{{\bm k}_2+}-\ee_{{\bm k}+}).
\eeqn
\end{widetext}
Making the same substitutions as for the third-order term and performing the impurity spin averages we obtain $\displaystyle\bigg(\frac{1}{\tau_m}\bigg)^{(4)}$ as will show below.

The expansion is straightforwardly continued to fifth order and above.
The Born approximation is sufficient for treating $U_{imp}$ and $U_{ep}$. In order to capture the relevant many-body Kondo physics, the Kondo scattering term must be evaluated in all orders in $U_m$. Whereas the number of terms increases with each order, the leading divergent terms, which are logarithmic in temperature, can be easily identified, and their contribution to the resistivity will be seen to form a straightforward geometric progression. We focus on these terms in this work, which is equivalent to summing the \textit{parquet diagrams}.

The kinetic equation is
\beqn
(\mathcal{J}_{ep} + \mathcal{J}_{imp} + \mathcal{J}_m)(f_{{\bm k}n}) = \frac{e{\bm E}}{\hbar}\cdot\pd{f_{0{\bm k}n}}{\bm k}.
\eeqn
We focus on the electron band $n = +$. The kinetic equation is readily solved by making the Ansatz $\displaystyle f_{E{\bk+}} \propto (e\mathbf{E}\cdot\hat{\mathbf{k}}/\hbar)(\partial f_{0{\bm k}+}/\partial k)$ \cite{Culcer_TI_Kineq_PRB10}. Then the full scattering term can be reduced to $\mathcal{J}(f_{\gamma}) = f_\gamma/\tau$, where $\frac{1}{\tau} = \frac{1}{\tau_{ep}} + \frac{1}{\tau_{imp}} + \frac{1}{\tau_m}$. For electron-phonon scattering in the Born approximation, we find
\beqn
\frac{1}{\tau_{ep}} = \frac{\pi^5 C^2}{15 \rho Ak_F^2 v_{ph}^5} \bigg(\frac{k_BT}{\hbar}\bigg)^4.
\eeqn
For scalar impurity scattering, from Eq.\ (\ref{Born}) [$\gamma = \theta' - \theta$]
\beqn
\frac{1}{\tau_{imp}} = \frac{n_i k_F}{4\hbar A}\int \frac{d\gamma}{2\pi}|\bar{V}^C_{\bk\bk'}|^2\sin^2\gamma.
\eeqn
For magnetic impurities, in the Born approximation,
\beqn
\bigg(\frac{1}{\tau_m}\bigg)^{(2)} = \frac{7 n_m S(S+1) J^2 k_F}{24A\hbar}.
\eeqn
To order $J^3$ [Eq. (\ref{JKondo})], we find the Kondo scattering term
\beqn
\bigg(\frac{1}{\tau_m}\bigg)^{(3)} = \frac{7n_mS(S+1) J^3k_F^2}{24A^2\hbar\pi} \ln\bigg|\frac{\ee_{k_F}}{k_BT}\bigg|.
\eeqn
We have retained the leading divergent terms, logarithmic in temperature, responsible for the Kondo physics, omitting a temperature-independent term in $J^3$. The exact result is found by summing all terms in the perturbation theory. The fourth order term yields
\begin{equation}
\bigg(\frac{1}{\tau_m}\bigg)^{(4)} = \frac{7n_m S(S+1)J^4k_F^3}{32\pi^2 A^3\hbar}\ln^2\bigg|\frac{\ee_F}{k_BT}\bigg|
\end{equation}
We sum all leading terms in $1/\tau_m$ exactly, obtaining
\beqn
\frac{1}{\tau_m} = \frac{7\pi n_m S(S+1)J^2 \rho_F}{12\hbar}\frac{1}{(1+J\rho_F\ln\big|\frac{k_BT}{\ee_F}\big|)^2},
\eeqn
where $\rho_F = \frac{k_F}{2\pi A}$ is the density of states at the Fermi energy. This diverges at the Kondo temperature
\beqn
T_K = \frac{\ee_F}{k_B}\exp\bigg( -\frac{1}{J\rho_F} \bigg).
\eeqn
This result is valid for arbitrary impurity spin. For the case of dilute alloys, the formalism outlined above reproduces results found previously \cite{Nagaoka_Kondo_PR65, Hamann_Kondo_EqMtn_PRB67, Silverstein_Duke_PhysRev.161.470, AbrikosovMigdal_Kondo_JLTP70} by summing the transition matrix or equivalent alternative methods. One simplification available in dilute alloys is the assumption of a short-range impurity potential, which enables one to sum the transition matrix exactly and use the optical theorem to deduce the transition rate immediately. This is an accurate approximation in transport because in metals for a short-range potential the transport lifetime is identical to the Bloch lifetime. In TI, due to the presence of terms prohibiting backscattering, the transport lifetime is always different from the Bloch lifetime, and approximating the transport lifetime using the optical theorem is not accurate.

The similarity in the expression for $T_K$ for TI and metals has been pointed out previously, and attributed to a peculiarity of the Rashba Hamiltonian of TI surface states, which allows the problem to be mapped to the pseudogap Anderson model \cite{Zitko_TI_Kondo_PRB10}. We expect the mathematical similarity of the two problems to extend beyond the Rashba model, since time reversal breaking by the magnetic impurities enables backscattering and eliminates topological protection in the many-body Kondo scattering terms.

The solution of the kinetic equation is $f_{{\bm k}+} = \frac{e{\bm E}\cdot\hat{\bm k}\tau}{2\hbar} \pd{f_0}{k}$. From this, the full resistivity is
\begin{widetext}
\begin{equation}
\rho_{xx} (T) = \frac{8\pi\hbar^2}{Ae^2k_F}\bigg[\frac{1}{\tau_{imp}} + \frac{7\pi n_m S(S+1)}{12\hbar}\frac{\rho_F J^2}{[1+J\rho_F\ln\bigg|\frac{k_BT}{\ee_F}\bigg|]^2} + \frac{\pi^5 C^2}{15 \rho Ak_F^2 v_{ph}^5} \bigg(\frac{k_BT}{\hbar}\bigg)^4 \bigg].
\end{equation}
\end{widetext}

We discuss the range of Kondo temperatures achievable in TI. It is reasonable to assume $J \approx 100$meV nm$^2$, based on figures reported in Ref.\ \onlinecite{Yu_TI_QuantAHE_Science10}, where the coupling constant is given as $ J_{eff} \approx 2$ eV. This value is normalized per Bi$_2$Se$_3$ unit, yielding $J = J_{eff} x V_{cell}$, where $x$ is the doping level (typically around 5$\%$) and $V_{cell}$ the unit cell volume. These values are also comparable to those in ferromagnetic semiconductors \cite{Jungwirth_FmgSmc_RMP06}. Using $A = 4.1$ eV$\AA$ \cite{Culcer_TI_PhysE12} and assuming $k_F \approx 10^8$m$^{-1}$, which corresponds to a doping density of $10^{11}$cm$^{-2}$ a typical number in quasi-2D systems, yields $\varepsilon_F \approx 500$ K. The exponent, however, makes $T_K$ negligible. The resistivity minimum, which also depends on the details of phonon scattering, is found by setting $\pd{}{T}\rho_{xx}(T) = 0$, which yields 0.9 K (it is 1.1 K at the rather high density of $10^{13}$ cm$^{-2}$). The location of the resistivity minimum is $\sim n_m^{1/4}$ (as for a 2D metal), as opposed to $\sim n_m^{1/5}$ for a 3D metal. Considering that these parameters are optimistic, we conclude that under realistic experimental conditions, the temperature dependence of the resistivity stems primarily from phonons.

\section{Discussion}
\label{Sec:Disc}

The Kondo temperature quantifies the tendency towards Kondo singlet formation between local moments and the Fermi sea. The small $T_K$ reflects the difficulty for a local moment to polarize the surface states and create a Kondo screening cloud, which stems from the strong coupling of the conduction electron spin to the momentum. The two energy scales appearing in $T_K$ are $Ak_F$ and $Jk_F^2$, the spin-orbit and exchange energies respectively. The Kondo effect in TI reflects the competition between these two mechanisms: the in-plane spin-orbit effective field prevents the impurity spin from polarising the conduction electrons. Hence, in TI spin-orbit coupling competes against the Kondo interaction, in the same way that it competes against electron-electron interactions in suppressing Stoner instabilities \cite{Culcer_TI_PhysE12}. Although spin-flip scattering is allowed, it is extremely unlikely. In general spin-momentum locking does \textit{not} guarantee the suppression of backscattering. For this reason, we expect our findings to persist when warping terms, important in Bi$_2$Te$_3$ \cite{Hasan_TI_RMP10}, are added to $H_{TI}$. This can be handled using the present formalism, which is general and can be applied to spin-orbit Hamiltonians of the form $H_{SO} = (\hbar/2) \, {\bm \sigma} \cdot {\bm \Omega}_{\bm k}$.

The Kondo temperature could be increased by increasing the density, though the bulk conduction band provides a stringent limit, given the small bulk gaps in current TI of the order of 0.3 eV. In the opposite limit, as the Dirac point is approached $T_K$ decreases exponentially since $\rho_F$ vanishes, yet transport near the Dirac point is diffusive and dominated by puddles \cite{Adam_TI_Tnsp_PRB12, Culcer_TI_Kineq_PRB10}. Our findings cannot be extrapolated to that regime.

The suppression of $T_K$ is contrasted with semiconductors and metals with strong spin-orbit coupling \cite{Zarea_PRL12, Vekhter_PRB12}, which differ from TI in several ways: TI have a single Fermi surface, spin-orbit coupling is strong, there is \textit{no} spin precession (indeed no spin-Hall effect) and no interband scattering. The simplest models of spin-orbit coupled semiconductors and metals consist of two bands, spin-orbit coupling is weak compared to the kinetic energy, the spin precesses in an effective field set by the spin-orbit interaction, interband scattering at the Fermi energy is just as important as intraband scattering, and the density of states in 2D is a constant. The much weaker spin-orbit in these conductors does not impede the formation of a Kondo screening cloud, and may under certain circumstances favor it \cite{Zarea_PRL12, Vekhter_PRB12}.

\section{Summary}
\label{Sec:Summary}

In summary, we have studied the Kondo effect in TI in the metallic regime, showing that the Kondo temperature is strongly suppressed by spin-momentum locking. The temperature dependence of the resistivity is due to phonon scattering. A natural extension of this work would be TI thin films, in which tunneling is possible between different TI surfaces.

\acknowledgments

We are grateful to O.~P.~Sushkov, R.~Winkler, Bruce Normand, S.~Das Sarma, Ion Garate, A.~H.~MacDonald, N.~P.~Butch, Yongqing Li and Tommy Li for stimulating discussions and to Z.~Fang for clarifying the situation in magnetic TI. The authors were in part supported by the Chinese Academy of Sciences.


\end{document}